\documentclass[prd,twocolumn]{revtex4-2}
\usepackage{mathrsfs}
\usepackage{amsmath}
\usepackage{amssymb}
\usepackage{graphicx}
\usepackage{braket}
\usepackage{hyperref}
\usepackage{color}
\usepackage{dsfont}
\usepackage[usenames,dvipsnames]{xcolor}

\usepackage{soul}
\usepackage[whole]{bxcjkjatype}

\hypersetup{
    colorlinks,
    citecolor=blue,
    linkcolor=blue,
    urlcolor=blue
}

\begin{document}
\title{Relativistic entropy production for quantum field in cavity}
\author{Yoshihiko Hasegawa}
\email{hasegawa@biom.t.u-tokyo.ac.jp}

\affiliation{Department of Information and Communication Engineering, Graduate
School of Information Science and Technology, The University of Tokyo,
Tokyo 113-8656, Japan}
\date{\today}
\begin{abstract}
A nonuniformly accelerated quantum field in a cavity undergoes the
coordinate transformation of annihilation and creation operators,
known as the Bogoliubov transformation. This study considers
the entropy production of a quantum field in a cavity induced by the
Bogoliubov transformation. By classifying the modes in the cavity into the system
and environment, we obtain the lower bound of the entropy production, defined as the sum of the von Neumann entropy in the system and the heat dissipated to the environment. This lower bound
 represents the refined second law of thermodynamics for a quantum
field in a cavity and can be interpreted as the Landauer principle,
which yields the thermodynamic cost of changing information contained
within the system. Moreover, it provides an upper bound for the quantum mutual
information to quantify the extent of the information scrambling
in the cavity due to acceleration. 
\end{abstract}
\maketitle

\section{Introduction}

The validity of classical mechanics is challenged in processes involving massive spatial scale as relativistic effects become nonnegligible at this scale.
For instance, clocks in satellites of the global positioning
system (GPS) tick faster than those on the ground owing to speed
and gravity, and thus, the accuracy of GPS deteriorates without
considering the relativistic effects. 
About half a century ago, the discoveries of Hawking radiation \cite{Hawking:1974:Radiation} and Unruh effect \cite{Crispino:2008:UnruhReview} revealed that incorporating relativistic effects in quantum mechanics can yield surprising phenomena that cannot be realized through conventional quantum mechanics. However, more recently, certain concepts in quantum information, such as entanglement, have been recognized as crucial in relativistic settings 
\cite{Mann:2012:RQI,Peres:2004:RQIReview}.
Currently, an entanglement distribution was performed between a
satellite and receivers on the ground located 1200 km apart
\cite{Yin:2017:Entanglement}. 
Against this background, relativistic quantum information has garnered considerable research attention, focusing on the relativistic effects of quantum information, e.g., Unruh \cite{Crispino:2008:UnruhReview} and dynamical Casimir effects \cite{Dodonov:2010:DCEReview}, using the quantum field theory \cite{Fuentes:2005:AliceBH,Downes:2011:EntCavity,Bruschi:2012:AlphaCentauri,MartinMartinez:2013:RQIComp,Bruschi:2013:RQComp,Ahmadi:2014:RMetrology}.

Quantum thermodynamics is an extension of the stochastic thermodynamics
operating at the mesoscopic scale to quantum microscopic scale. It
generalizes the concept of stochastic work, heat, and entropy to the
quantum domain, and accordingly, several thermodynamic relations, e.g., Jarzynski
equality \cite{Jarzynski:1997:Equality}, fluctuation theorem \cite{Evans:1993:FT,Gallavotti:1995:Ensemble,Crooks:1999:CFT},
and thermodynamic uncertainty relation \cite{Barato:2015:UncRel,Horowitz:2019:TURReview},
have been demonstrated to hold in the quantum domain as well \cite{Mukamel:2003:QJE,Funo:2018:QFT,Hasegawa:2020:TUROQS}.
Recently, the thermodynamic quantities and relations have been further generalized
to quantum field theory. References~\cite{Bartolotta:2018:JEQFT}
and \cite{Ortega:2019:WorkQFT} derived the Jarzynski equality for
quantum field theory in a flat spacetime using two-point and indirect
measurements, respectively. Regarding the quantum thermodynamics for quantum
field theory in a curved spacetime, Ref.~\cite{Liu:2016:QFTQT} investigated
the work exerted by the expanding universe. Moreover, Ref.~\cite{Bruschi:2020:RQT}
considered several quantities to formulate the quantum thermodynamics of a quantum field in an accelerating cavity.

Herein, we study the entropy production \cite{Landi:2021:EPReview}
in a quantum field confined in a cavity that undergoes acceleration.
In thermodynamics, the entropy production quantifies the extent of
irreversibility of the system as well as the thermodynamic cost of thermal machines. The nonnegativity of the
entropy production is a signature of the second law of thermodynamics
and directly implies the Landauer principle \cite{Landauer:1961:LP},
which formulates the relation between information, quantified by entropy,
and dissipated heat. Despite its significance in quantum thermodynamics,
the entropy production induced by acceleration has not been investigated
thus far. We consider a quantum field experiencing arbitrary acceleration in a cavity to induce a coordinate transformation referred
to as the Bogoliubov transformation. By defining the entropy production
based on the sum of entropy and dissipated heat in a quantum field, we can derive
the lower bound of the entropy production, corresponding to a refinement of
the second law of thermodynamics for the quantum field in a cavity, resulting in the
Landauer principle. Moreover, using the obtained inequality, an upper bound of the quantum mutual information can be obtained to
quantify the extent of information scrambling caused by acceleration.

\section{Methods}

We consider a $(1+1)$ dimensional Minkowski space \cite{Friis:2013:CavityRQI}.
Suppose that a cavity  of length $L>0$ contains a massless scalar field. The confined quantum field model has
been extensively employed in relativistic quantum information. The
scalar quantum field satisfies the Klein-Goldon equation in a curved spacetime
\cite{Mukhanov:2007:QGravityBook}. The field $\Phi$ admits the mode
expansion with respect to $\{\phi_{n}\}_{n=1}^{\infty}$, expressed as 
\begin{equation}
\hat{\Phi}=\sum_{n}(\hat{a}_{n}\phi_{n}+\hat{a}_{n}^{\dagger}\phi_{n}^{*}),\label{eq:mode_expansion1}
\end{equation}
where $\hat{a}_{n}$ and $\hat{a}_{n}^{\dagger}$ denote the annihilation
and creation operators, respectively, which satisfy the canonical commutation
relation $[\hat{a}_{n},\hat{a}_{m}^{\dagger}]=\delta_{n,m}$ and
$[\hat{a}_{n},\hat{a}_{m}]=[\hat{a}_{n}^{\dagger},\hat{a}_{m}^{\dagger}]=0$. 
A coordinate transformation between
different observers, induced by acceleration can
be modeled based on the Bogoliubov transformation, which transforms the modes
$\{\phi_{n}\}$ in the original coordinate to modes $\{\tilde{\phi}_{n}\}$
in another coordinate. Moreover, the field $\hat{\Phi}$ can be expanded
by $\{\tilde{\phi}_{n}\}_{n=1}^{\infty}$ as well:
\begin{equation}
\hat{\Phi}=\sum_{n}(\hat{b}_{n}\tilde{\phi}_{n}+\hat{b}_{n}^{\dagger}\tilde{\phi}_{n}^{*}),\label{eq:mode_expansion2}
\end{equation}
where $\hat{b}_{n}$ and $\hat{b}_{n}^{\dagger}$ represent distinct annihilation
and creation operators, respectively, satisfying the canonical commutation
relation $[\hat{b}_{n},\hat{b}_{m}^{\dagger}]=\delta_{n,m}$ and $[\hat{b}_{n},\hat{b}_{m}]=[\hat{b}_{n}^{\dagger},\hat{b}_{m}^{\dagger}]=0$.
$\hat{a}_{n}$ and $\hat{b}_{n}$ are related via 
\begin{equation}
\hat{b}_{m}=\sum_{n}(A_{mn}\hat{a}_{n}+B_{mn}\hat{a}_{n}^{\dagger}),\label{eq:Bogoliubov_trans}
\end{equation}
where $A_{mn}$ and $B_{mn}$ are the Bogoliubov coefficients.
The matrices $A=\{A_{mn}\}$ and $B=\{B_{mn}\}$ should satisfy the
Bogoliubov identities $AA^{\dagger}-BB^{\dagger}=\mathds{1}$ and
$AB^{\top}-BA^{\top}=0$, where $\mathds{1}$ is the identity matrix.
For all $n\in\{1,2,\cdots\}$, the annihilation operator $\hat{a}_{n}$ defines the vacuum state
$\ket{0}$ via $\hat{a}_{n}\ket{0}=0$.
Therefore, the vacuum state $\ket{0}$ represents an eigenstate with a vanishing
eigenvalue of the annihilation operator. One of the most prominent
properties of the Bogoliubov transformation is that the vacuum states
of different coordinates are not generally consistent with each other. Indeed, the vacuum
state $\ket{\tilde{0}}$ for $\hat{b}_{n}$ can be expressed as $\hat{b}_{n}\ket{\tilde{0}}=0$
for all $n$, which does not agree with $\ket{0}$ in general. Therefore,
the vacuum state of a coordinate may be populated with particles with
respect to another coordinate.

\begin{figure}
\includegraphics[width=8cm]{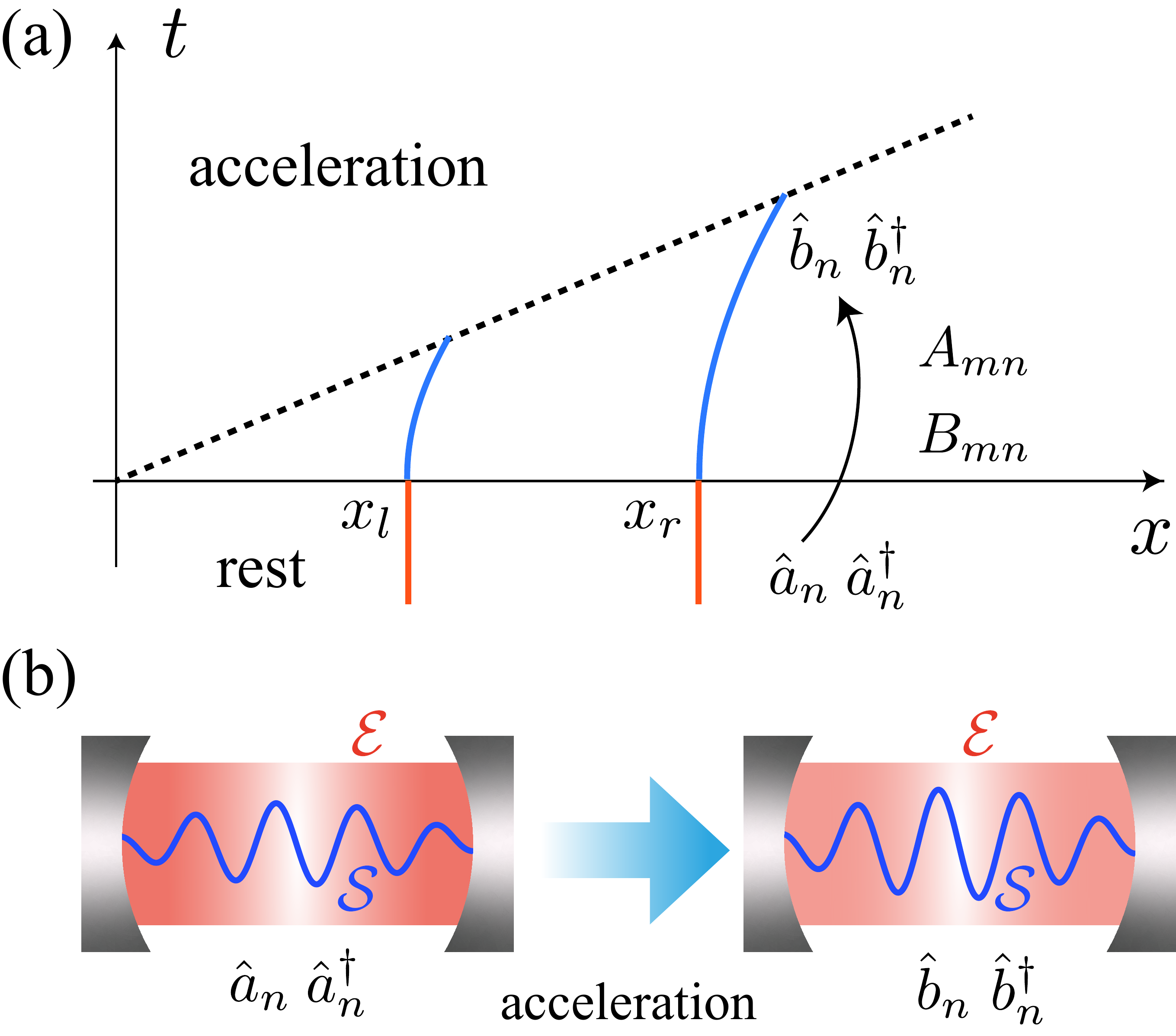} \caption{ Illustration of relativistic quantum field in a cavity. (a) Example
of cavity trajectory in Minkovski space. $x_{l}$ and $x_{r}$
are boundaries of the cavity, where $x_{r}-x_{l}=L>0$. The cavity
is inertial for $t<0$ and starts to accelerate for $t\ge0$. Due
to the acceleration, the annihilation and creation operators undergoes
the Bogoliubov transformation specified by $A_{mn}$ and $B_{mn}$.
(b) Separation of system and environment. Modes specified
by $\mathcal{S}$ correspond to the system and the remaining modes
$\mathcal{E}$ relate to the environment. \label{fig:model_ponch}}
\end{figure}

Suppose that the cavity is in an inertial frame at the initial state.
In relativistic quantum information, we are typically interested
in only one or two modes in the cavity, whereas the remaining modes are
regarded as the environment \cite{Bruschi:2012:AlphaCentauri,Bruschi:2020:RQT}.
Therefore, following Ref.~\cite{Bruschi:2020:RQT}, we divide 
all the modes into the system $\mathcal{S}$ and environment $\mathcal{E}$.
A set of modes in the system $\mathcal{S}$ is defined as $\mathcal{S}=\{n_{s_{1}},n_{s_{2}},\ldots,n_{s_{K}}\}$,
where $n_{s_{i}}\in\{1,2,\cdots\}$ denotes an index of the system mode
satisfying $n_{s_{i}}\ne n_{s_{j}}$ for $i\ne j$, $K$ indicates the
number of modes of the system, and a set of modes in the environment
$\mathcal{E}$ comprises the remaining modes, i.e., $\mathcal{E}=\{n\in\{1,2,\cdots\}|n\notin\mathcal{S}\}$.
Thus, a set of all the cavity modes can be expressed as $\mathcal{C}=\mathcal{E}\cup\mathcal{S}=\{1,2,\cdots\}$.

As the quantum field is an infinite-dimensional system, the thermodynamic quantities cannot be easily calculated. Thus, we employ the Gaussian state
formalism \cite{Ferraro:2005:GaussianBook,Wang:2007:GSReview,Adesso:2014:CVQIReview} that is widely used in relativistic quantum information to compute
thermodynamic quantities. Let $\hat{\xi}\equiv[\hat{a}_{1},\hat{a}_{2},...,\hat{a}_{1}^{\dagger},\hat{a}_{2}^{\dagger},...]^{\top}$.
Subsequently, the covariance matrix $\sigma=\{\sigma_{mn}\}$ can be defined as
\begin{equation}
\sigma_{mn}\equiv\braket{\hat{\xi}_{m}\hat{\xi}_{n}^{\dagger}+\hat{\xi}_{n}^{\dagger}\hat{\xi}_{m}}-2\braket{\hat{\xi}_{m}}\braket{\hat{\xi}_{n}^{\dagger}},\label{eq:cov_sigma_def}
\end{equation}
where $\braket{\bullet}$ denotes the expectation value. Let $\sigma^{i}$
and $\sigma^{f}$ denote the covariance matrices before and after the
coordinate transformation, respectively (hereinafter, variables
with superscripts $i$ and $f$ indicate the stated aspect). In particular, the covariance matrix after Bogoliubov transformation can be expressed as  
\begin{equation}
\sigma^{f}=\mathfrak{S}\sigma^{i}\mathfrak{S}^{\dagger},
\label{eq:sigma_evolution}
\end{equation}
where $\mathfrak{S}$ denotes a complex symplectic transformation, specified by the
Bogoliubov matrices $A$ and $B$: 
\begin{equation}
\mathfrak{S}=\left[\begin{array}{cc}
A & B\\
B^{*} & A^{*}
\end{array}\right].\label{eq:S_def}
\end{equation}
The thermodynamic quantities of interest can be represented by the covariance
matrix $\sigma$.

\section{Results}

Let us consider a massless quantum field in a cavity that is initially in an
inertial frame. According to the Klein--Goldon
equation, the field mode $\phi_{n}$ ($n\in\{1,2,\cdots\}$) can be expressed as 
\begin{equation}
\phi_{n}(t,x)=\frac{1}{\sqrt{n\pi}}\sin\left[\omega_{n}(x-x_{l})\right]e^{-i\omega_{n}t},\label{eq:phi_inertial_def}
\end{equation}
where we select a comoving frame $(t,x)$ with $x_{l}$ and $x_{r}$ as the cavity boundaries for any $t$ with $x_{r}-x_{l}=L>0$
(refer to Fig.~\ref{fig:model_ponch}(a)), and the Dirichlet boundary
condition is imposed at the boundaries. The Hamiltonian operators
of the system and environment are respectively defined as 
\begin{align}
\hat{H}_{\mathrm{sys}}\equiv\sum_{n\in\mathcal{S}}\omega_{n}\hat{a}_{n}^{\dagger}\hat{a}_{n},\label{eq:Hsys_def}\\
\hat{H}_{\mathrm{env}}\equiv\sum_{n\in\mathcal{E}}\omega_{n}\hat{a}_{n}^{\dagger}\hat{a}_{n},\label{eq:Henv_def}
\end{align}
where $\omega_{n}\equiv\pi n/L$ denotes the angular frequency of the field
mode. 
Let $\beta_\mathrm{sys}$ and $\beta_\mathrm{env}$ be the inverse temperature of the system and environment, respectively.
We define the thermal states of the system and environment as follows:
\begin{align}
\rho_{\mathrm{sys}}^{\mathrm{th}}(\beta_{\mathrm{sys}})&\equiv\frac{1}{Z_{\mathrm{sys}}(\beta_{\mathrm{sys}})}e^{-\beta_{\mathrm{sys}}\hat{H}_{\mathrm{sys}}},\label{eq:rho_th_sys_def}\\
\rho_{\mathrm{env}}^{\mathrm{th}}(\beta_{\mathrm{env}})&\equiv\frac{1}{Z_{\mathrm{env}}(\beta_{\mathrm{env}})}e^{-\beta_{\mathrm{env}}\hat{H}_{\mathrm{env}}},
\label{eq:rho_th_env_def}
\end{align}
where $Z_{\mathrm{sys}}(\beta_{\mathrm{sys}})$ and $Z_{\mathrm{env}}(\beta_{\mathrm{env}})$ are the partition functions defined as
\begin{align}
Z_{\mathrm{sys}}(\beta_{\mathrm{env}})&\equiv\mathrm{Tr}_{\mathrm{sys}}[e^{-\beta_{\mathrm{env}}\hat{H}_{\mathrm{sys}}}],\label{eq:Z_sys_def}\\
Z_{\mathrm{env}}(\beta_{\mathrm{env}})&\equiv\mathrm{Tr}_{\mathrm{env}}[e^{-\beta_{\mathrm{env}}\hat{H}_{\mathrm{env}}}].\label{eq:Z_env_def}
\end{align}
The initial density operator of the cavity is assumed to be $\rho_{\mathrm{cav}}^{i}=\rho_{\mathrm{sys}}^{i}\otimes\rho_{\mathrm{env}}^{i}$,
where $\rho_{\mathrm{sys}}^{i}$ and $\rho_{\mathrm{env}}^{i}$ are
given by $\rho_{\mathrm{sys}}^{i}=\rho_{\mathrm{sys}}^{\mathrm{th}}(\beta_{\mathrm{sys}}^{i})$
and $\rho_{\mathrm{env}}^{i}=\rho_{\mathrm{env}}^{\mathrm{th}}(\beta_{\mathrm{env}}^{i})$
with $\beta_{\mathrm{sys}}^{i}$ and $\beta_{\mathrm{env}}^{i}$ representing
the initial inverse temperature of the system and environment, respectively.
Let $\eta_{n}\equiv\mathrm{coth}(\beta_{\mathrm{sys}}^i\omega_{n}/2)$
and $\nu_{n}\equiv\mathrm{coth}(\beta_{\mathrm{env}}^i\omega_{n}/2)$
be the symplectic eigenvalues of the system and environment, respectively.
The initial Gaussian states of the system and environment are 
respectively defined as
\begin{align}
\sigma_{\mathrm{sys}}^{i} & =\mathrm{diag}([\eta_{n}]_{n\in\mathcal{S}},[\eta_{n}]_{n\in\mathcal{S}}),\label{eq:sigma_s_def}\\
\sigma_{\mathrm{env}}^{i} & =\mathrm{diag}([\nu_{n}]_{n\in\mathcal{E}},[\nu_{n}]_{n\in\mathcal{E}}).\label{eq:sigma_def}
\end{align}
Based on symplectic eigenvalues, the mean energy of the initial environment
state can be derived as (refer to Appendix~\ref{sec:QFT_quantities})
\begin{equation}
\mathrm{Tr}_{\mathrm{env}}\left[\rho_{\mathrm{env}}^{i}\hat{H}_{\mathrm{env}}\right]=\sum_{n\in\mathcal{E}}\omega_{n}\frac{\sigma_{nn}^{i}-1}{2}.\label{eq:mean_energy_C}
\end{equation}
Furthermore, the von Neumann entropy can be defined as 
\begin{align}
S_{\mathrm{sys}}(\rho_{\mathrm{sys}})\equiv-\mathrm{Tr}_{\mathrm{sys}}[\rho_{\mathrm{sys}}\ln\rho_{\mathrm{sys}}].
\label{eq:Ssys_def}
\end{align}
The von Neumann entropy is defined similarly for the environment $S_{\mathrm{env}}$ and cavity $S_{\mathrm{cav}}$. In the covariance matrix formalism, the von Neumann entropy
can be represented by a symplectic eigenvalue of the system \cite{Serafini:2017:QCV}:
\begin{equation}
S_{\mathrm{sys}}=\sum_{n\in\mathcal{S}}\left\{ \mathfrak{s}_{+}\left(\eta_{n}\right)-\mathfrak{s}_{-}\left(\eta_{n}\right)\right\} ,\label{eq:vnEntropy_def}
\end{equation}
where $\mathfrak{s}_{\pm}(x)\equiv\left\{ (x\pm1)/2\right\} \ln\left\{ (x\pm1)/2\right\} $.

After preparing the initial state, the cavity undergoes arbitrary
acceleration. An example of the trajectory for $t>0$ is exhibited in Fig.~\ref{fig:model_ponch}(a), where the cavity starts to accelerate
satisfying the rigidity of the cavity. The coordinate transformation
induced by the acceleration is modeled by the Bogoliubov transformation.
Let $\hat{\zeta}\equiv[\hat{b}_{1},\hat{b}_{2},...,\hat{b}_{1}^{\dagger},\hat{b}_{2}^{\dagger},...]^{\top}$.
The Bogoliubov transformation on operators $\hat{\xi}$ can be
unitarily implemented as follows \cite{Simon:1988:Symplectic,Arvind:1995:Symplectic,Ferraro:2005:GaussianBook}:
\begin{equation}
\hat{\zeta}=\mathfrak{S}\hat{\xi}=\hat{U}^{\dagger}(\mathfrak{S})\hat{\xi}\hat{U}(\mathfrak{S}),
\label{eq:xi_transformation}
\end{equation}
where $\mathfrak{S}$ denotes a symplectic transformation defined as Eq.~\eqref{eq:S_def}
and $\hat{U}(\mathfrak{S})$ denotes a unitary operator satisfying $\hat{U}(\mathfrak{S}_{1}\mathfrak{S}_{2})=\hat{U}(\mathfrak{S}_{1})\hat{U}(\mathfrak{S}_{2})$.
Equation~\eqref{eq:xi_transformation} is the Heisenberg picture
of the creation and annihilation operators. Therefore, in the Schr{\"o}dinger
picture, the density operator of the entire cavity evolves unitarily
as $\rho_{\mathrm{cav}}^{f}=\hat{U}(\mathfrak{S})\rho_{\mathrm{cav}}^{i}\hat{U}^{\dagger}(\mathfrak{S})$.
In the quantum thermodynamics, the dissipated heat is often defined by
the energy difference in the environment between the final and initial
states:
\begin{align}
\Delta Q&\equiv\mathrm{Tr}_{\mathrm{env}}[\rho_{\mathrm{env}}^{f}\hat{H}_{\mathrm{env}}]-\mathrm{Tr}_{\mathrm{env}}[\rho_{\mathrm{env}}^{i}\hat{H}_{\mathrm{env}}]\nonumber\\&=\sum_{n\in\mathcal{E}}\omega_{n}\frac{\sigma_{nn}^{f}-\sigma_{nn}^{i}}{2},
 \label{eq:heat_def}
\end{align}
where $\rho_{\mathrm{env}}^{f}\equiv\mathrm{Tr}_{\mathrm{sys}}[\rho_{\mathrm{cav}}^{f}]=\mathrm{Tr}_{\mathrm{sys}}[\hat{U}\rho_{\mathrm{cav}}^{i}\hat{U}^{\dagger}]$
denotes the final density operator of the environment. The entropy difference
between the initial and final states can be expressed as 
\begin{equation}
\Delta S_{\mathrm{sys}}\equiv S_{\mathrm{sys}}(\rho_{\mathrm{sys}}^{f})-S_{\mathrm{sys}}(\rho_{\mathrm{sys}}^{i}),
\label{eq:entropy_difference}
\end{equation}
where the covariance matrix representation of $S_{\mathrm{sys}}$
follows that in Eq.~\eqref{eq:vnEntropy_def}.

Subsequently, we define the entropy production for a quantum field in a cavity.
The entropy production plays fundamental roles in stochastic and quantum
thermodynamics, which can be defined following several manner
\cite{Seifert:2012:FTReview,Landi:2021:EPReview}. For instance, in
stochastic thermodynamics, the entropy production can be quantified by the probability
ratio between the forward and backward processes, or it may be defined
by a total entropy that includes both the system and environment. In the quantum
domain, owing to the high degree of freedom in modeling, the entropy production can be defined following several mechanisms \cite{Landi:2021:EPReview}. Here,
we define the entropy production as
\begin{align}
\Sigma\equiv\beta_{\mathrm{env}}^{i}\Delta Q+\Delta S_{\mathrm{sys}},
    \label{eq:entropy_production_def}
\end{align}
which denotes the sum of the dissipated heat {[}Eq.~\eqref{eq:heat_def}{]}
and the von Neumann entropy of the system {[}Eq.~\eqref{eq:entropy_difference}{]}.
As the modes in the cavity undergo Bogoliubov transformation,
the density operator of the entire cavity $\rho_{\mathrm{cav}}$ evolves
via the corresponding unitary operator {[}Eq.~\eqref{eq:xi_transformation}{]}.
Therefore, from Refs.~\cite{Esposito:2010:EntProd,Reed:2014:Landauer},
the following relation holds (refer to Appendix~\ref{sec:Landauer}): 
\begin{align}
\Sigma=I+D(\rho_{\mathrm{env}}^{f}||\rho_{\mathrm{env}}^{i})\ge D(\rho_{\mathrm{env}}^{f}||\rho_{\mathrm{env}}^{i})\ge0,
 \label{eq:Landauer}
\end{align}
where $D(\rho_{\mathrm{env}}^{f}||\rho_{\mathrm{env}}^{i})$ and $I$
are the quantum relative entropy and the quantum mutual information,
respectively, defined by 
\begin{align}
D(\rho_{\mathrm{env}}^{f}||\rho_{\mathrm{env}}^{i})&\equiv\mathrm{Tr}_{\mathrm{env}}\left[\rho_{\mathrm{env}}^{f}\ln\rho_{\mathrm{env}}^{f}\right]-\mathrm{Tr}_{\mathrm{env}}\left[\rho_{\mathrm{env}}^{f}\ln\rho_{\mathrm{env}}^{i}\right],\label{eq:QrelEnt_def}\\
I&\equiv S_{\mathrm{sys}}(\rho_{\mathrm{sys}}^{f})+S_{\mathrm{env}}(\rho_{\mathrm{env}}^{f})-S_{\mathrm{cav}}(\rho_{\mathrm{cav}}^{f}).\label{eq:QMutInfo_def}
\end{align}
The quantum relative entropy and quantum mutual information are nonnegative \cite{Nielsen:2011:QuantumInfoBook} (nonnegativity of quantum mutual information used in the second line of Eq.~\eqref{eq:Landauer}).
As expressed in Eq.~\eqref{eq:Landauer}, the entropy production is nonnegative, $\Sigma \ge 0$, under a coordinate transformation induced by the acceleration, which is a second law of thermodynamics for a quantum field in a cavity.

Prior to delving into deeper analysis of the entropy production $\Sigma$, we explain certain aspects. 
In this study, we defined heat as the variation in the energy of the environment. 
Heat is induced by the interaction between the system and environment.
Although the exchange of energy between the system and environment is identified as heat in classical thermodynamics, 
the distinction of heat from work in the quantum setting is often nontrivial. 
Prior studies have attempted to define heat and work in quantum thermodynamics, which can be primarily classified into two fundamental definitions, namely, heat-first and work-first definitions \cite{Elouard:2018:QTraj}. 
In the heat-first definition, heat is defined as the variation in the energy of the environment \cite{Hekking:2013:QJump}.
This definition is commonly applied to the two-point measurement scheme \cite{Campisi:2011:QuantumFT}, where projective measurement with respect to the environmental energy eigenbasis is performed at the beginning and end of the process, and the heat is defined as the difference between them. 
In the second category, work is defined first, which was initially proposed in Ref~\cite{Alicki:1979:HeatEngine}. 
This study employed the heat-first definition.

The first term in Eq.~\eqref{eq:entropy_production_def} represents the increase in environmental entropy, which is true for the ideal environment that maintains its equilibrium and with constant temperature during the evolution. 
However, the definition of temperature in general nonequilibrium states persists to be a challenge \cite{CasasVazquez:2003:NoneqTemperature}. 
Therefore, in the standard quantum thermodynamics, the entropy production in Eq.~\eqref{eq:entropy_production_def} is defined as the entropy production for an arbitrary environment \cite{Landi:2021:EPReview}. 
This is because $\Sigma$ defined in Eq.~\eqref{eq:entropy_production_def} is positive in all cases, consistent with the most essential requirement for entropy production. 
Moreover, the definition of Eq.~\eqref{eq:entropy_production_def} is consistent with that of classical thermodynamics. 
Note that the entropy production in Eq.~\eqref{eq:entropy_production_def} exhibits the operational meaning in case of considering a fluctuation theorem \cite{Manzano:2018:EPPRX}, which is the fundamental equality in nonequilibrium thermodynamics.

This study focuses on heat but not on work, which constitutes the counterpart quantity in thermodynamics. 
As the unitary $\hat{U}(\mathfrak{S})$ in Eq.~\eqref{eq:xi_transformation} implicitly includes the contribution of work,
the variation in energy of the system is not equal to the heat, $\Delta H_\mathrm{sys} + \Delta Q \ne 0$ \cite{Landi:2021:EPReview}, where $\Delta H_\mathrm{sys}$ represents the variations in the system energy:
\begin{align}
    \Delta H_{\mathrm{sys}}&\equiv\mathrm{Tr}_{\mathrm{sys}}\left[\rho_{\mathrm{sys}}^{f}\hat{H}_{\mathrm{sys}}\right]-\mathrm{Tr}_{\mathrm{sys}}\left[\rho_{\mathrm{sys}}^{i}\hat{H}_{\mathrm{sys}}\right]\nonumber\\&=\sum_{n\in\mathcal{S}}\frac{\sigma_{nn}^{f}-\sigma_{nn}^{i}}{2}.
    \label{eq:Delta_Hsys_def}
\end{align}
According to the first law, the exerted work can be defined as the difference between $\Delta \hat{H}_\mathrm{sys}$ and $\Delta Q$:
\begin{align}
    \Delta W\equiv\Delta H_{\mathrm{sys}}+\Delta Q.
    \label{eq:work_W_def}
\end{align}
Based on the work defined in Eq.~\eqref{eq:work_W_def}, the entropy production of Eq.~\eqref{eq:entropy_production_def} can be represented as 
\begin{align}
    \Sigma=\beta_{\mathrm{env}}^{i}\left(\Delta W-\Delta F\right),
    \label{eq:EP_as_freeenergy}
\end{align}
where $\Delta F\equiv F(\rho_{\mathrm{sys}}^{f})-F(\rho_{\mathrm{sys}}^{i})$ represents the variation in free energy, and $F(\rho_{\mathrm{sys}})$ indicates the free energy defined as
\begin{align}
    F(\rho_{\mathrm{sys}})\equiv\mathrm{Tr}_{\mathrm{sys}}[\hat{H}_{\mathrm{sys}}\rho_{\mathrm{sys}}]-\frac{1}{\beta_{\mathrm{env}}^{i}}S_{\mathrm{sys}}(\rho_{\mathrm{sys}}).
    \label{eq:free_energy_def}
\end{align}
The second law
$\Sigma \ge 0$ yields $\Delta W\ge\Delta F$, which states that the work exerted on the system is greater than or equal to the free energy difference, which is consistent with another classical definition of entropy production.

\begin{figure}
\includegraphics[width=8cm]{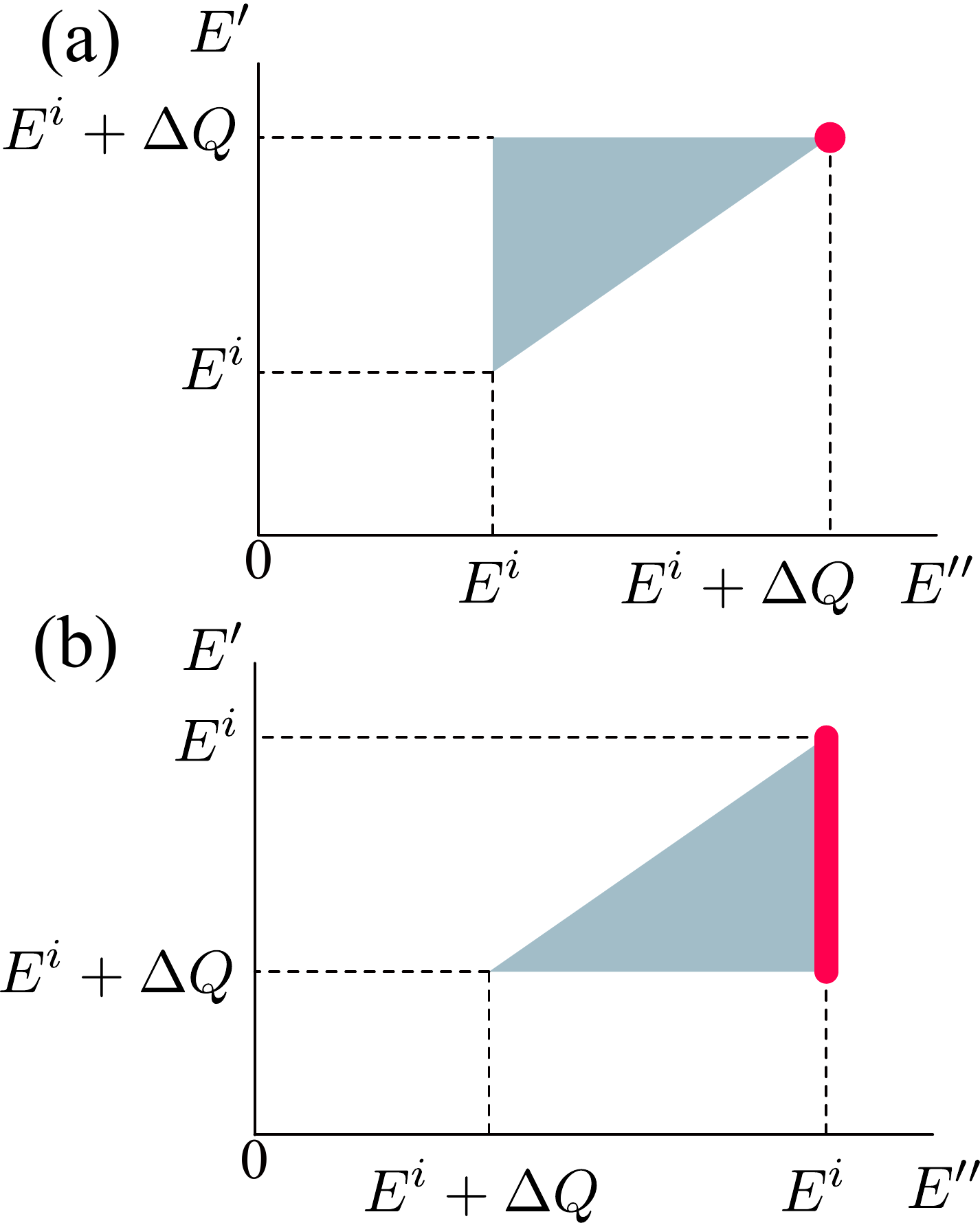} \caption{ Integral regions of Eq.~\eqref{eq:D_ineq} for (a) $\Delta Q>0$
and (b) $\Delta Q<0$. Pink regions denote values of $E^{\prime\prime}$
and $E^{\prime}$ at which $\mathrm{Var}_{\beta}[\hat{H}_{\mathrm{env}}]$
gives the maximum. \label{fig:integral_region}}
\end{figure}

\begin{figure}
\includegraphics[width=7cm]{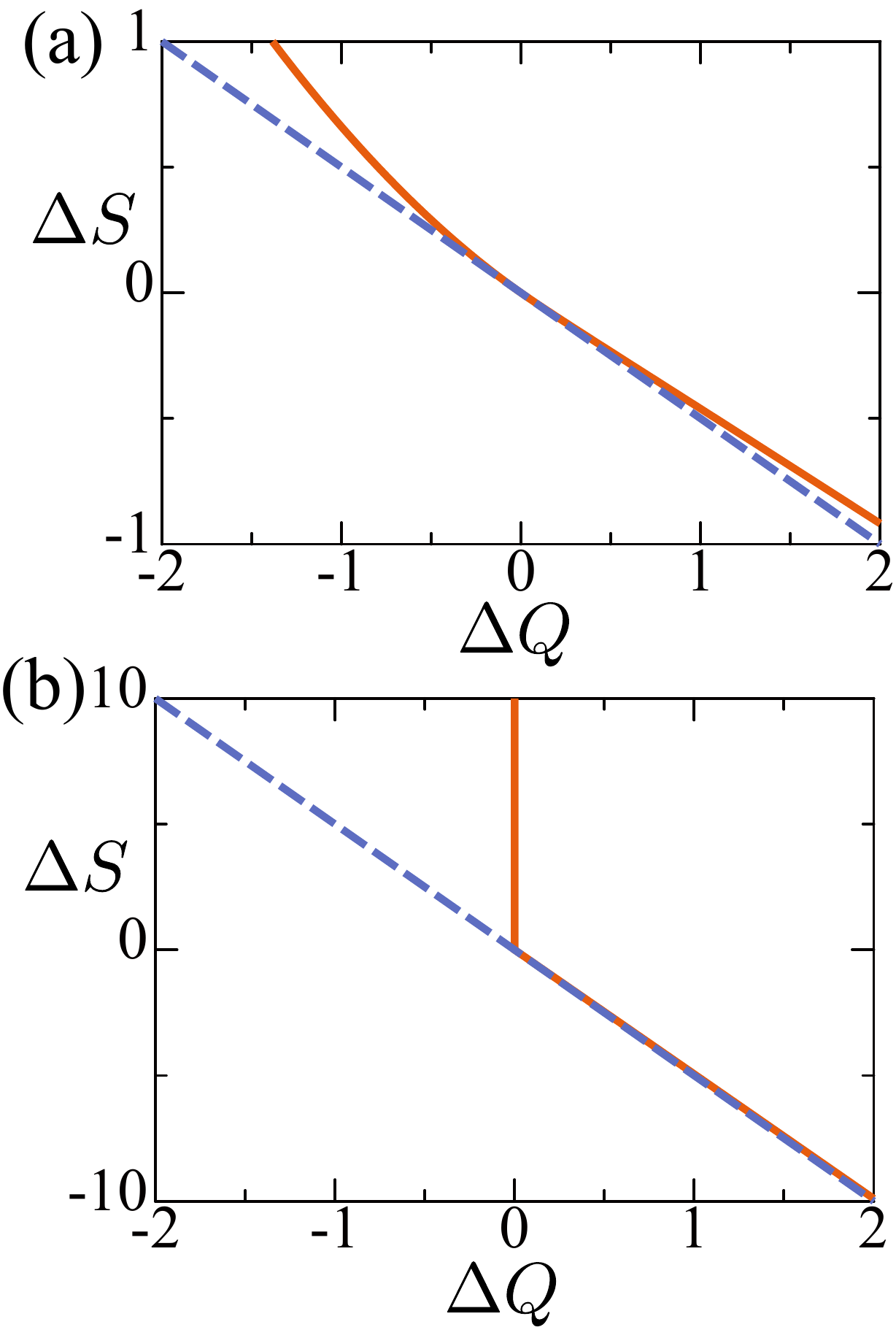} \caption{ Region plot of lower bound {[}Eq.~\eqref{eq:main_result}{]}
for cavity with (a) $\beta_{\mathrm{env}}^{i}=0.5$ and (b) $\beta_{\mathrm{env}}^{i}=5.0$.
The solid line represents Eq.~\eqref{eq:main_result} and the dashed
line denotes $\Delta S_{\mathrm{sys}}\ge-\beta_{\mathrm{env}}^{i}\Delta Q$.
The regions above the solid lines are physically feasible, whereas the other settings
are $\mathcal{S}=\{1\}$ and $L=1$. \label{fig:Landauer_bound_plot}}
\end{figure}

We can refine Eq.~\eqref{eq:Landauer} by using the fact that the
environment comprises a bosonic quantum field. To calculate the lower
bound of $D(\rho_{\mathrm{env}}^{f}||\rho_{\mathrm{env}}^{i})$, we
follow Ref.~\cite{Reed:2014:Landauer}. According to the Pythagoras relation
\cite{Reed:2014:Landauer} that can be proved by simple calculations, $D(\rho_{\mathrm{env}}^{f}||\rho_{\mathrm{env}}^{i})$
is bounded from below by 
\begin{align}
D(\rho_{\mathrm{env}}^{f}||\rho_{\mathrm{env}}^{i})&=D(\rho_{\mathrm{env}}^{f}||\rho_{\mathrm{env}}^{\mathrm{th},f})+D(\rho_{\mathrm{env}}^{\mathrm{th},f}||\rho_{\mathrm{env}}^{i})\nonumber\\&\ge D(\rho_{\mathrm{env}}^{\mathrm{th},f}||\rho_{\mathrm{env}}^{i}),\label{eq:Pitagoras}
\end{align}
where $\rho_{\mathrm{env}}^{\mathrm{th},f}$ denotes a thermal state that
yields the same energy as $\rho_{\mathrm{env}}^{f}$. Let $E(\beta_\mathrm{env})$
be the mean energy of the environment with respect to a thermal state
with the inverse temperature $\beta_\mathrm{env}$: 
\begin{equation}
E(\beta_{\mathrm{env}})\equiv\mathrm{Tr}_{\mathrm{env}}\left[\rho_{\mathrm{env}}^{\mathrm{th}}(\beta_{\mathrm{env}})\hat{H}_{\mathrm{env}}\right].
\label{eq:EE_def}
\end{equation}
Thereafter, we obtain 
\begin{equation}
E^{i}+\Delta Q=\mathrm{Tr}_{\mathrm{env}}[\rho_{\mathrm{env}}^{f}\hat{H}_{\mathrm{env}}]=\mathrm{Tr}_{\mathrm{env}}[\rho_{\mathrm{env}}^{\mathrm{th},f}\hat{H}_{\mathrm{env}}],
\label{eq:E_plus_DQ}
\end{equation}
where $E^{i}\equiv E(\beta_{\mathrm{env}}^{i})=\mathrm{Tr}_{\mathrm{env}}[\rho_{\mathrm{env}}^{i}\hat{H}_{\mathrm{env}}]$.
As $(d/d\beta_{\mathrm{env}})E(\beta_{\mathrm{env}})<0$, $\beta_{\mathrm{env}}^{\mathrm{th},f}$
satisfies $E(\beta_{\mathrm{env}}^{\mathrm{th},f})=E^{i}+\Delta Q$ and
can be uniquely specified given $\Delta Q$. Therefore, given
$\Delta Q$, $\rho_{\mathrm{env}}^{\mathrm{th},f}=\rho_{\mathrm{env}}^{\mathrm{th}}(\beta_{\mathrm{env}}^{\mathrm{th},f})$
can be uniquely identified, indicating that $D(\rho_{\mathrm{env}}^{\mathrm{th},f}||\rho_{\mathrm{env}}^{i})$
in Eq.~\eqref{eq:Pitagoras} can be calculated with $\Delta Q$.
The relative entropy admits the following expression: 
\begin{equation}
D(\rho_{\mathrm{env}}^{\mathrm{th},f}||\rho_{\mathrm{env}}^{i})=\beta_{\mathrm{env}}^{i}\Delta Q-\left[S_{\mathrm{env}}(\rho_{\mathrm{env}}^{\mathrm{th},f})-S_{\mathrm{env}}(\rho_{\mathrm{env}}^{i})\right].\label{eq:D_decomposition}
\end{equation}
Based on Ref.~\cite{Reed:2014:Landauer}, $D(\rho_{\mathrm{env}}^{\mathrm{th},f}||\rho_{\mathrm{env}}^{i})$ in Eq.~\eqref{eq:D_decomposition}
can be rewritten as (refer to Appendix~\ref{sec:improved_derivation}):
\begin{align}
D(\rho_{\mathrm{env}}^{\mathrm{th},f}||\rho_{\mathrm{env}}^{i})=\int_{E^{i}}^{E^{i}+\Delta Q}dE^{\prime}\int_{E^{i}}^{E^{\prime}}\frac{dE^{\prime\prime}}{\mathrm{Var}_{\beta_{\mathrm{env}}(E^{\prime\prime})}[\hat{H}_{\mathrm{env}}]},\label{eq:D_ineq}
\end{align}
where $\mathrm{Var}_{\beta_{\mathrm{env}}}[\hat{H}_{\mathrm{env}}]$ denotes the variance
of $\hat{H}_{\mathrm{env}}$ with respect to the thermal state of the
environment with inverse temperature $\beta_\mathrm{env}$: 
\begin{align}
\mathrm{Var}_{\beta_{\mathrm{env}}}[\hat{H}_{\mathrm{env}}]=\frac{1}{4}\sum_{n\in\mathcal{E}}\omega_{n}^{2}\mathrm{csch}\left(\frac{\beta_{\mathrm{env}}\omega_{n}}{2}\right)^{2},\label{eq:varH_cavity}
\end{align}
where $\beta_\mathrm{env}(E)$ denotes the inverse temperature $\beta_\mathrm{env}$, satisfying $E=\mathrm{Tr}_{\mathrm{env}}\left[\rho_{\mathrm{env}}^{\mathrm{th}}(\beta_{\mathrm{env}})\hat{H}_{\mathrm{env}}\right]$,
which can be uniquely identified because $E(\beta_\mathrm{env})$ is a monotonically
decreasing function. 
Thereafter, we calculate the maximum value of $\mathrm{Var}_{\beta_\mathrm{env}}[\hat{H}_{\mathrm{env}}]$
within the integral domain $\int_{E^{i}}^{E^{i}+\Delta Q}dE^{\prime}\int_{E^{i}}^{E^{\prime}}dE^{\prime\prime}$.
Specifically, we need to separately consider the two cases of $\Delta Q>0$ and $\Delta Q<0$. As $\beta_\mathrm{env}(E)$ is a monotonically decreasing
function of $E$, $\mathrm{Var}_{\beta_\mathrm{env}}[\hat{H}_{\mathrm{env}}]$
provides the maximum value for $\Delta Q>0$ and $\Delta Q<0$ at the
points indicated in Fig.~\ref{fig:integral_region}(a) and (b), respectively.
Considering the maximum, we derive a refined version of the second law for the accelerated cavities as follows: 
\begin{equation}
\Delta S_{\mathrm{sys}}+\beta_{\mathrm{env}}^{i}\Delta Q\ge\begin{cases}
{\displaystyle \frac{(\Delta Q)^{2}}{2\mathrm{Var}_{\beta_{\mathrm{env}}^{i}}[\hat{H}_{\mathrm{env}}]}} & \Delta Q\le0\\
{\displaystyle \frac{(\Delta Q)^{2}}{2\mathrm{Var}_{\beta_{\mathrm{env}}(E^{i}+\Delta Q)}[\hat{H}_{\mathrm{env}}]}} & \Delta Q>0
\end{cases},\label{eq:main_result}
\end{equation}
which forms the main result of this research. In Eq.~\eqref{eq:main_result},
the variance term for $\Delta Q<0$ does not rely on $\Delta Q$,
unlike $\Delta Q>0$. Equation~\eqref{eq:main_result}
holds for an arbitrary Bogoliubov transformation, indicating that
Eq.~\eqref{eq:main_result} should hold for any acceleration undergone
by the cavity. Equation~\eqref{eq:main_result} is a refined version of
the second law for the quantum field in the cavity. Although the above
calculation follows Ref.~\cite{Reed:2014:Landauer}, the lower bound
of Eq.~\eqref{eq:main_result} differs from that reported in Ref.~\cite{Reed:2014:Landauer}
because the cavity is an infinite-dimensional system, whereas Ref.~\cite{Reed:2014:Landauer}
concerns finite-dimensional systems. 

Although Eq.~\eqref{eq:main_result}
represents the statement for entropy production, it can be regarded as
a statement between the variation in information in the system and the energy dissipated
in the environment, i.e., Eq.~\eqref{eq:main_result} can
be identified as the Landauer principle for a quantum field in the
cavity undergoing acceleration. The Landauer principle concerns
the entropy decrease in the system, quantified by $-\Delta S_{\mathrm{sys}}$,
and yields the lower bound of the heat dissipation to realize the
entropy decrease. Equation~\eqref{eq:main_result}
is plotted in Fig.~\ref{fig:Landauer_bound_plot} with the solid lines for two inverse temperature settings in (a) $\beta_{\mathrm{env}}^{i}=0.5$
and (b) $\beta_{\mathrm{env}}^{i}=5.0$ (explicit parameters are stated 
in the caption). The regions above the solid
lines denote the feasible regions predicted by Eq.~\eqref{eq:main_result}.
In Fig.~\ref{fig:Landauer_bound_plot}, the dashed lines represent
the lower bound of $\Sigma=\Delta S_{\mathrm{sys}}+\beta_{\mathrm{env}}^{i}\Delta Q\ge0$,
which represents the naive second law. As observed, the area of the negative
heat region diminishes with the temperature. More importantly, Eq.~\eqref{eq:main_result} is tighter than
the naive second law under high temperature.

Another consequence of Eq.~\eqref{eq:main_result} pertains to its relation
with information scrambling \cite{Swingle:2018:OTOC,Larkin:1969:OTOC}.
Generally, the extent of scrambling is quantified by out-of-order correlators.
As proposed earlier, the extent of scrambling can alternatively be quantified by quantum mutual information \cite{Touil:2021:Scrambling}.
The cavity undergoing a nonuniform acceleration can be identified
as a process of information spreading. Suppose that the cavity contains
modes only in $\mathcal{S}$ and the remaining modes are vacant. After
acceleration, the other remaining modes are populated
because of the Bogoliubov transformation, which is a reminiscent of the
scrambling process. The mutual information $I$ is bounded from above
by $I\le\Delta S_{\mathrm{sys}}+\beta_{\mathrm{env}}^{i}\Delta Q-D(\rho_{\mathrm{env}}^{\mathrm{th},f}||\rho_{\mathrm{env}}^{i})$.
Based on Eq.~\eqref{eq:main_result}, we obtain 
\begin{equation}
I\le\begin{cases}
\beta_{\mathrm{env}}^{i}\Delta Q+\Delta S_{\mathrm{sys}}-{\displaystyle \frac{(\Delta Q)^{2}}{2\mathrm{Var}_{\beta_{\mathrm{env}}^{i}}[\hat{H}_{\mathrm{env}}]}} & \Delta Q\le0\\
\beta_{\mathrm{env}}^{i}\Delta Q+\Delta S_{\mathrm{sys}}-{\displaystyle \frac{(\Delta Q)^{2}}{2\mathrm{Var}_{\beta_{\mathrm{env}}(E^{i}+\Delta Q)}[\hat{H}_{\mathrm{env}}]}} & \Delta Q>0
\end{cases},\label{eq:I_upper_bound}
\end{equation}
which indicates that the degree of scrambling induced by the acceleration of the cavity can be upper-bounded by the entropy $\Delta S_{\mathrm{sys}}$ and dissipated heat $\Delta Q$ in the cavity.

\section{Conclusion}

In this paper, we obtained the lower bound for the entropy production
of a quantum field in a cavity undergoing acceleration. First,
the cavity mode of interest was regarded as the system and the 
remaining modes as the environment. Thereafter, the entropy production
was defined as the sum of the von Neumann entropy of the system and the dissipated heat. The nonnegativity of the entropy production is a signature
of the second law and provides the statement of the Landauer
principle for the accelerated cavity.

\begin{acknowledgments}
This work was supported by JSPS KAKENHI Grant Numbers JP19K12153 and JP22H03659.
\end{acknowledgments}

\appendix

\section{Quantities of quantum field\label{sec:QFT_quantities}}
For readers' convenience, 
we review the quantities of quantum fields in a general setting. 
Let us consider the thermal state of the Hamiltonian $\hat{H}=\sum_{n}\omega_{n}\hat{a}_{n}^{\dagger}\hat{a}_{n}$.
The density operator can be expressed as 
\begin{equation}
\rho^{\mathrm{th}}\equiv\frac{1}{Z(\beta)}e^{-\beta\hat{H}},\label{eq:rho_th_def}
\end{equation}
where $\beta$ denotes the inverse temperature and $Z(\beta)\equiv\mathrm{Tr}[e^{-\beta\hat{H}}]$.
The number operator $\hat{\mathfrak{n}}_{n}\equiv\hat{a}_{n}^{\dagger}\hat{a}_{n}$
admits the eigendecomposition 
\begin{equation}
\hat{\mathfrak{n}}_{n}\ket{\mathfrak{n}_{n}}=\mathfrak{n}_{n}\ket{\mathfrak{n}_{n}},\label{eq:number_op_decomposition}
\end{equation}
which implies that $\hat{\mathfrak{n}}_{n}$ can be represented as
$\hat{\mathfrak{n}}_{n}=\sum_{\mathfrak{n}_{n}}{\mathfrak{n}}_{n}\ket{\mathfrak{n}_{n}}\bra{\mathfrak{n}_{n}}$.
Based on this representation, the terms in $\rho^{\mathrm{th}}$ can be expressed as 
\begin{equation}
e^{-\beta\hat{H}}=\prod_{n=1}^{\infty}e^{-\beta\omega_{n}\hat{\mathfrak{n}}_{n}}=\prod_{n=1}^{\infty}\sum_{\mathfrak{n}_{n}}e^{-\beta\omega_{n}\mathfrak{n}_{n}}\ket{\mathfrak{n}_{n}}\bra{\mathfrak{n}_{n}},\label{eq:thermal_state_1}
\end{equation}
and 
\begin{equation}
\mathrm{Tr}[e^{-\beta\hat{H}}]=\prod_{n}\left(\sum_{\mathfrak{n}_{n}}e^{-\beta\omega_{n}\mathfrak{n}_{n}}\right)=\prod_{n}\frac{e^{\beta\omega_{n}}}{e^{\beta\omega_{n}}-1}.\label{eq:eBH_trace}
\end{equation}
Let us consider the expectation of $\hat{a}_{n}^{\dagger}\hat{a}_{n}$
with respect to the thermal state $\rho^{\mathrm{th}}$: 
\begin{equation}
\braket{\hat{a}_{n}^{\dagger}\hat{a}_{n}}=\mathrm{Tr}\left[\hat{\mathfrak{n}}_{n}\rho^{\mathrm{th}}\right]=\frac{\sum_{\mathfrak{n}_{n}}\mathfrak{n}_{n}e^{-\beta\omega_{n}\mathfrak{n}_{n}}}{\sum_{\mathfrak{n}_{n}}e^{-\beta\omega_{n}\mathfrak{n}_{n}}}=\frac{1}{e^{\beta\omega_{n}}-1}.\label{eq:number_op_ev}
\end{equation}
As $\ensuremath{\sigma_{nn}=\braket{\hat{a}_{n}^{\dagger}\hat{a}_{n}+\hat{a}_{n}\hat{a}_{n}^{\dagger}}}=2\braket{\hat{a}_{n}^{\dagger}\hat{a}_{n}}+1$,
the covariance matrix becomes 
\begin{equation}
\sigma_{nn}=2\braket{\hat{a}_{n}^{\dagger}\hat{a}_{n}}+1=\mathrm{coth}\left(\frac{\beta\omega_{n}}{2}\right).\label{eq:sigma_nn}
\end{equation}
According to Eq.~\eqref{eq:number_op_ev}, the mean of $\hat{H}$ can be derived as
\begin{equation}
\mathrm{Tr}[\rho^{\mathrm{th}}\hat{H}]=\sum_{n}\frac{\omega_{n}}{e^{\beta\omega_{n}}-1}.\label{eq:mean_energy}
\end{equation}
Similarly, the second moment of $\hat{H}$ can be evaluated as
\begin{align}
\mathrm{Tr}\left[\rho^{\mathrm{th}}\hat{H}^{2}\right] & =\sum_{n}\frac{(e^{\beta\omega_{n}}+1)\omega_{n}^{2}}{(e^{\beta\omega_{n}}-1)^{2}}\nonumber \\
 & +\sum_{n\ne m}\left(\frac{\omega_{n}}{e^{\beta\omega_{n}}-1}\right)\left(\frac{\omega_{m}}{e^{\beta\omega_{m}}-1}\right).\label{eq:H_second_moment}
\end{align}
Using Eqs.~\eqref{eq:mean_energy} and \eqref{eq:H_second_moment},
the variance of $\hat{H}$ is expressed as 
\begin{align}
\mathrm{Var}[\hat{H}] & =\mathrm{Tr}\left[\rho^{\mathrm{th}}\hat{H}^{2}\right]-\mathrm{Tr}\left[\rho^{\mathrm{th}}\hat{H}\right]^{2}\nonumber \\
 & =\sum_{n}\frac{e^{\beta\omega_{n}}\omega_{n}^{2}}{(e^{\beta\omega_{n}}-1)^{2}}\nonumber \\
 & =\frac{1}{4}\sum_{n}\omega_{n}^{2}\mathrm{csch}\left(\frac{\beta\omega_{n}}{2}\right)^{2},\label{eq:var_H}
\end{align}
where $\mathrm{csch}(x)\equiv1/\sinh(x)$. 

$E(\beta)$ is
defined in Eq.~\eqref{eq:EE_def} with a derivative of 
\begin{equation}
\frac{d}{d\beta}E(\beta)=-\sum_{n\in\mathcal{E}}\frac{e^{\beta\omega_{n}}\omega_{n}^{2}}{(e^{\beta\omega_{n}}-1)^{2}}<0.\label{eq:Ebeta_diff}
\end{equation}
The above equation establishes the definition of the inverse function $\beta(E)$.

\section{Derivation of Eq.~\eqref{eq:Landauer}\label{sec:Landauer}}

The derivation of Eq.~\eqref{eq:Landauer} is presented herein,
which has been proved in Refs.~\cite{Esposito:2010:EntProd,Reed:2014:Landauer}.
We will express the following relation:
\begin{equation}
\beta_{\mathrm{env}}^{i}\Delta Q+\Delta S_{\mathrm{sys}}=I+D(\rho_{\mathrm{env}}^{f}||\rho_{\mathrm{env}}^{i}).\label{eq:Landauer2}
\end{equation}
As discussed in the main text, the entire cavity undergoes a unitary
transformation: $\rho_{\mathrm{cav}}^{f}=U\rho_{\mathrm{cav}}^{i}U^{\dagger}$.
Therefore, the von Neumann entropy of the entire cavity is invariant
under the transformation: $S_{\mathrm{cav}}(\rho_{\mathrm{cav}}^{f})=S_{\mathrm{cav}}(\rho_{\mathrm{cav}}^{i})=S_{\mathrm{sys}}(\rho_{\mathrm{sys}}^{i})+S_{\mathrm{env}}(\rho_{\mathrm{env}}^{i})$,
where we considered that the state is initially in a product state
in the last equality.
Therefore, we obtain 
\begin{equation}
I+D(\rho_{\mathrm{env}}^{f}||\rho_{\mathrm{env}}^{i})=\Delta S+\mathrm{Tr}_{\mathrm{env}}\left[(\rho_{\mathrm{env}}^{i}-\rho_{\mathrm{env}}^{f})\ln\rho_{\mathrm{env}}^{i}\right].\label{eq:I_D_part}
\end{equation}
As $\rho_{\mathrm{env}}^{i}=Z_{\mathrm{env}}(\beta_{\mathrm{env}}^{i})^{-1}e^{-\beta_{\mathrm{env}}^{i}\hat{H}_{\mathrm{env}}}$,
the second term in Eq.~\eqref{eq:I_D_part} can be rewritten as 
\begin{align}
 & \mathrm{Tr}_{\mathrm{env}}\left[(\rho_{\mathrm{env}}^{i}-\rho_{\mathrm{env}}^{f})\ln\rho_{\mathrm{env}}^{i}\right]\nonumber\\
 & =\mathrm{Tr}_{\mathrm{env}}\left[(\rho_{\mathrm{env}}^{i}-\rho_{\mathrm{env}}^{f})(-\beta_{\mathrm{env}}^{i}\hat{H}_{\mathrm{env}})\right]\nonumber \\
 & =\beta_{\mathrm{env}}^{i}\Delta Q.\label{eq:trace_part}
\end{align}
Equation~\eqref{eq:Landauer} directly follows from Eqs.~\eqref{eq:I_D_part}
and \eqref{eq:trace_part}.

\section{Derivation of Eq.~\eqref{eq:D_ineq}\label{sec:improved_derivation}}

Herein, we derive Eq.~\eqref{eq:D_ineq} based on Refs.~\cite{Reed:2014:Landauer}:
\begin{align}
D(\rho_{\mathrm{env}}^{\mathrm{th},f}||\rho_{\mathrm{env}}^{i})&=\beta_{\mathrm{env}}^{i}\Delta Q-\int_{E^{i}}^{E^{i}+\Delta Q}\frac{dS_{\mathrm{env}}^{\mathrm{th}}(\beta_{\mathrm{env}}(E^{\prime}))}{dE^{\prime}}dE^{\prime}\nonumber\\&=\beta_{\mathrm{env}}^{i}\Delta Q-\int_{E^{i}}^{E^{i}+\Delta Q}\beta_{\mathrm{env}}(E^{\prime})dE^{\prime}\nonumber\\&=\int_{E^{i}}^{E^{i}+\Delta Q}\left(\beta_{\mathrm{env}}^{i}-\beta_{\mathrm{env}}(E^{\prime})\right)dE^{\prime}\nonumber\\&=\int_{E^{i}}^{E^{i}+\Delta Q}dE^{\prime}\int_{E^{\prime}}^{E^{i}}\frac{d\beta_{\mathrm{env}}(E^{\prime\prime})}{dE^{\prime\prime}}dE^{\prime\prime}\nonumber\\&=\int_{E^{i}}^{E^{i}+\Delta Q}dE^{\prime}\int_{E^{i}}^{E^{\prime}}\frac{dE^{\prime\prime}}{\mathrm{Var}_{\beta_{\mathrm{env}}(E^{\prime\prime})}[\hat{H}_{\mathrm{env}}]},\label{eq:D_ineq_app}
\end{align}
where $S_{\mathrm{env}}^{\mathrm{th}}(\beta_{\mathrm{env}})$ is defined as
\begin{equation}
S_{\mathrm{env}}^{\mathrm{th}}(\beta_{\mathrm{env}})\equiv-\mathrm{Tr}_{\mathrm{env}}\left[\rho_{\mathrm{env}}^{\mathrm{th}}(\beta_{\mathrm{env}})\ln\rho_{\mathrm{env}}^{\mathrm{th}}(\beta_{\mathrm{env}})\right].
\label{eq:Senvth_def}
\end{equation}
When calculating Eq.~\eqref{eq:D_ineq_app}, we used the following relations:
\begin{align}
    \frac{d\beta_{\mathrm{env}}}{dE}&=\frac{1}{dE/d\beta_{\mathrm{env}}}=-\frac{1}{\mathrm{Var}_{\beta_{\mathrm{env}}(E)}[\hat{H}_{\mathrm{env}}]},\label{eq:dbetadE}\\\frac{d\rho_{\mathrm{env}}^{\mathrm{th}}(\beta_{\mathrm{env}})}{d\beta_{\mathrm{env}}}&=-\hat{H}_{\mathrm{env}}\rho_{\mathrm{env}}^{\mathrm{th}}(\beta_{\mathrm{env}})\nonumber\\&+\rho_{\mathrm{env}}^{\mathrm{th}}(\beta_{\mathrm{env}})\mathrm{Tr}_{\mathrm{env}}\left[\rho_{\mathrm{env}}^{\mathrm{th}}(\beta_{\mathrm{env}})\hat{H}_{\mathrm{env}}\right],\label{eq:drhoenvthdbeta_calc}\\\frac{dS_{\mathrm{env}}^{\mathrm{th}}}{d\beta_{\mathrm{env}}}&=-\mathrm{Tr}_{\mathrm{env}}\left[\frac{d\rho_{\mathrm{env}}^{\mathrm{th}}(\beta_{\mathrm{env}})}{d\beta_{\mathrm{env}}}\ln\rho_{\mathrm{env}}^{\mathrm{th}}(\beta_{\mathrm{env}})\right]\nonumber\\&=-\beta_{\mathrm{env}}\mathrm{Var}_{\beta_{\mathrm{env}}}\left[\hat{H}_{\mathrm{env}}\right],\label{eq:dSdbeta_calc}\\\frac{dS_{\mathrm{env}}^{\mathrm{th}}(\beta_{\mathrm{env}}(E))}{dE}&=\frac{dS_{\mathrm{env}}^{\mathrm{th}}(\beta_{\mathrm{env}})}{d\beta_{\mathrm{env}}}\frac{d\beta_{\mathrm{env}}}{dE}=\beta_{\mathrm{env}}(E).\label{eq:dSdE}
\end{align}

\end{document}